\documentclass{sigchi-ext}
\pdfoutput=1
\usepackage[T1]{fontenc}
\usepackage{textcomp}
\usepackage[scaled=.92]{helvet} 
\usepackage{graphicx} 
\usepackage{balance}  
\usepackage{booktabs} 
\usepackage{ccicons}  
\usepackage{ragged2e} 

\usepackage{subfigure}
\usepackage{algorithm}
\usepackage{pseudocode}
\usepackage{algpseudocode}
\usepackage{amsmath}

\usepackage{endnotes}
\usepackage{commath}
\usepackage{mathtools}
\usepackage{ragged2e}

\def\plaintitle{Characterizing Smartphone Power Management in the Wild} 
\def\emptyauthor{}
\def\plainkeywords{Sate of Charge; Fast Charging; Battery Capacity.}

\title{Characterizing Smartphone Power Management in the Wild}

\numberofauthors{2}
\author{%
  \alignauthor{%
    \textbf{Mohammad A. Hoque}\\
    \affaddr{University of Helsinki} \\
    \affaddr{Helsinki, Finland} \\
    \email{mohammad.a.hoque@helsinki.fi} } \hspace{10mm}\alignauthor{%
      \textbf{Sasu Tarkoma}\\
   \affaddr{University of Helsinki} \\
    \affaddr{Helsinki, Finland} \\
    \email{sasu.tarkoma@helsinki.fi} }
    }

\definecolor{linkColor}{RGB}{6,125,233}
\hypersetup{%
  pdftitle={\plaintitle},
  pdfauthor={\emptyauthor},
  pdfkeywords={\plainkeywords},
  bookmarksnumbered,
  pdfstartview={FitH},
  colorlinks,
  citecolor=black,
  filecolor=black,
  linkcolor=black,
  urlcolor=linkColor,
  breaklinks=true,
}

\begin{document}

\maketitle

\RaggedRight{} 

\begin{abstract}
\justify
   For better reliability and prolonged battery life, it is important that users and vendors understand the quality of charging and the performance of smartphone batteries. Considering the diverse set of devices and user behavior it is a challenge. In this work, we analyze a large collection of battery analytics dataset collected from 30K devices of 1.5K unique smartphone models. We analyze their battery properties and state of charge while charging, and reveal the characteristics of different components of their power management systems: charging mechanisms, state of charge estimation techniques, and their battery properties. We explore diverse charging behavior of devices and their users.

\end{abstract}



\section{Introduction}
\justify
While modeling and optimizing the energy consumption of different applications and mobile systems have been the active research interest for a decade~\cite{6157576,hoquecsur2015}, the overall performance of the smartphone power management systems has not received  significant attention yet. The diverse set of smartphone models available today are powered with batteries of different capacity volumes and technologies, such as Lithium-Ion and Lithium-Polymer. They employ different charging mechanisms to charge their batteries and rely on different state of charge (SOC) estimation techniques. 

The growth of smartphone battery size has been linear with time. Charging large batteries with traditional charging techniques may take very long time. In addition, the context may not allow a user to charge long enough time. Therefore, it is necessary that the battery should be charged to some reasonable amount, e.g., 30-50\%, within a short amount of time. Consequently, users are increasingly relying on a number of Fast charging techniques from Qualcomm (Quick~\cite{qulcomm}) and Samsung (Fast). Nevertheless, the quality of charging plays an important role in the longevity of smartphone batteries. For example, if a battery is charged over the maximum battery voltage, the resulting chemical reactions may reduce the capacity significantly~\cite{Choi2002130} and increase the battery temperature beyond the safety limit. Understanding the inefficiency of the energy source, and other related  contributing factors can enable better optimization of the applications, systems, and more accurate power consumption modeling.

\marginpar{%
  \vspace{-21pc} \fbox{%
    \begin{minipage}{0.925\marginparwidth}
      \underline{\textbf{Findings and Contributions}} 
    {\small
    
      \vspace{2mm} 
      \underline{\textbf{First:}} Most of the devices apply Constant Current-Constant Voltage (CC-CV) charging. A number of devices of latest models use Fast charging. Quick charging utilizes higher voltage and constant current during the CC phase of charging, whereas Samsung's Fast charging employs pulse charging. 

      \vspace{1mm}\underline{\textbf{Second:}} From SOC updates, we compute the charging time curves, and derive the SOC estimation techniques used by the devices. The SOC update time of voltage-based fuel gauges fluctuates during the CC-phase of charging, whereas Coulomb counter-based devices provide updates at periodic intervals.
      
      \vspace{1mm} \underline{\textbf{Third:}} From battery voltage, we estimate  that 85\% devices had  their battery capacity reduced by 1-10\%. We further demonstrate that the battery health information provided by the system may not be appropriate.
      
      \vspace{1mm} \underline{\textbf{Fourth:}} Both devices and users contribute to inefficient charging of the batteries. Quick charging charges a battery to over voltage. There are two forms of inefficient charging by the users; charging and actively using devices at the same time and longer over night charging. 
      }
    \end{minipage}}\label{sec:sidebar} }

Given the number of smart mobile devices available on the market, it is not feasible to investigate their charging and battery properties, and the performance of the charging methods on batteries in a laboratory environment.  Although there are studies on users' charging behavior~\cite{Ferreira:2011,Banerjee:2007}, it is not well understood how this battery and charging information could be presented in more meaningful ways to the users and mobile vendors other than just the battery level.

In this article, we explore a large battery analytic dataset comprising various battery sensor information from 30K devices of 1.5K unique smartphone models collected by the Carat~\cite{Oliner:2013} application. We explore their battery voltage behavior, charging rate and charging time, and demonstrate how these properties can be used to expose the characteristics of their power management systems. We identify their charging mechanisms, SOC estimation techniques and battery properties, and the distribution of these properties among the devices. To the best of our knowledge such comprehensive study on a large smartphone battery dataset has not been presented earlier. Our findings and contributions are listed in the sidebar.

The rest of the paper is organized as follows. Next section provides an overview on smartphone's power management system and describes the crowdsourced dataset. The subsequent  sections explore the dataset and identify the characteristics of various power management techniques used by the smartphones and properties of smartphone batteries while charging. Before concluding the paper, we also discuss user behavior in charging their smartphones. 

\section{Smartphone Power Management \& {Dataset}}
\label{sec:two}

The charger and three different ICs, a fuel gauge, a charging controller, and a protection IC,  manage the charging of a mobile device. The charging controller is hosted in the device and the protection IC resides in the battery. The fuel gauge functionality may be distributed between the device and the battery.  The fuel gauge determines the runtime battery capacity, i.e., SOC or battery level, using open circuit voltage, coulomb counter, or a combined mechanism of these two~\cite{Rezvanizaniani2014110}.  It senses battery voltage,  temperature, and charge or discharge current to/from the battery pack. At the same time, it also provides feedback to the charging IC. The charging controller applies the charging algorithm, such as CC-CV, and uses the fuel gauge provided information to control the charging current, voltage, and to terminate the charging. Finally, the protection IC protects the battery from over voltage or current from the device. 

\subsection{Dataset and Pre-processing}
The Android Battery Manager collects charging and battery information from the fuel gauge (see Table~\ref{tab:charger_battery_info}) and broadcasts as events. 
Carat collects information from mobile devices as samples with a broadcast receiver. A sample structure  can be defined as $S = (t, (a_{1} : v_{1}), (a_{2} : v_{2}), (a_{3} : v_{3})...(a_{n} : v_{n}))$, where $t$ is the epoch timestamp of a SOC update event and $(a_{i} : v_{i})$ are the attribute and value pairs. From all the information collected in a sample, we  consider the timestamp,  SOC, battery voltage, battery health, battery temperature,  charging status, charger type, and the screen status attributes.

\begin{margintable}[1pc]
  \begin{minipage}{\marginparwidth}
  \vspace{22mm}
 
   \begin{tabular}{|p{12mm}|p{24mm}|}
      \hline
       \textbf{Attribute}  & \textbf{Value}\\\hline
       charging status & plugged/\break unplugged\\\hline
       charger & ac main / usb\\\hline
       battery voltage & 4.2 V\\\hline
       battery temp &  29$^\circ$C \\\hline
       battery health & good,overheat, over~voltage\\\hline
       battery level & 0.99 (99\%)\\\hline
    \end{tabular}
       \caption{Charging and battery  information of Android devices.}
       \vspace{1mm}
     \label{tab:charger_battery_info}
     
       \vspace{7mm}
  \end{minipage}
\end{margintable}

\marginpar{%
  \fbox{%
    \begin{minipage}{0.925\marginparwidth}
    
    \begin{equation}
\label{eq:one}
C = \frac{36\times \Delta SOC}{{t}_{2}-{t}_{1}}
\end{equation}

        \end{minipage}}\label{sec:sidebar2} }

\begin{algorithm}[t]
\caption{Smartphone Battery  Analytics}\label{alg:cevent}
\begin{algorithmic}[1]
\Procedure{ChargingEvents~}{ rdd[reducedSample] }
\State userSamps$\leftarrow$ groupById (reducedSample)
\For{\textbf{each} \texttt{user} $\in$ \texttt{userSamps.\_1}}
    \State filSamp $\leftarrow$ Filter (``ac'', ``soff'', userSamp)
    \State srtSamp $\leftarrow$ SortByTime (filSamp)
    \State samPairs $\leftarrow$ Pair (srtSamp)
    \State chargEvent = 1
    \For{\textbf{each} \texttt{pair} $\in$ \texttt{samPairs}}
        \State $\Delta$SOC = onePair.\_1.soc - onePair.\_2.soc
        \State $\Delta$t = onePair.\_2.time - onePair.\_1.time
        \State $C$  = $\frac{36\times\Delta SOC}{\Delta t}$
    
        \If{\texttt{($C<=0.03$)}}
            \State Label (chargEvent, $\Delta$t, C, onePair.\_1)
            \State chargEvent += 1
            \State $\Delta$t = 0
            \State $C$ = 0
            \State Label (chargEvent, $\Delta$t, C, onePair.\_2)
        
        \Else
            \State Label (chargEvent, 0, 0, onePair.\_1)
            \State Label (chargEvent, $\Delta$t, C, onePair.\_2)
        \EndIf
    \EndFor
\EndFor
\EndProcedure
\end{algorithmic}
\label{alg:one}
\end{algorithm}

\vspace{-2mm}
\subsection{Charging Events}

We analyze a subset of Carat dataset collected over ten months of size more than 200GB, in Spark platform~\cite{spark}. From the reduced charging samples, we need to construct SOC vs. battery voltage and SOC vs. charging rate curves. In order to generate such curves, it is essential that all the samples of a curve belong to the same charging event. We generate charging events for every user as described in Algorithm \ref{alg:one}.  The reduced samples in a Resilient Distributed Dataset allows distributed computation on the samples in a cluster of 7 machines each having 8 CPU cores and 30GB of RAM.  

\vspace{-1mm}
First, the reduced samples are grouped according to the user ID. We next sort the samples of a user according to timestamp and construct pairs of two consecutive samples. From these pairs,  we compute one percent charging time and corresponding C\footnote{\footnotesize{A battery with 2000 mAh capacity will be charged with 2000 mA current at 1.0C rate and it will take 1 hour to complete the charging.}} rate. If the charging rate is 1C, then a mobile device spends 36 s to charge one percent. As a result, the charging rate used to charge 1\% of the battery can be presented as \eqref{eq:one}, where $t_{1},~t_{2}$ are the timestamps of two consecutive samples and $\Delta$SOC is the difference of the battery levels reported in those samples.

Ideally, a charging event begins by connecting a device with the charger and ends when the device is disconnected from the charger. Constructing charging events in this way is difficult from the dataset, as a user may turn on/off the phone while charging and turn on when the battery is charged to a reasonable capacity. The charging algorithms terminate charging when the charging rate is 0.07C and a mobile device spends 514 seconds maximum to charge one percent. However, the tablets may take even longer time during the CV phase of charging and therefore, we consider 0.03C as the terminating charging rate as stated in the algorithm. Finally,  we label the samples with incremental numeric charging events, and update them with C rate and one percent charging time.

\begin{table}[t]
\begin{center}
  {\small
    \begin{tabular}{|p{17mm}|p{12mm}|p{9mm}|p{11mm}|p{13mm}|}
      \hline
       \textbf{Battery\break Health} &\textbf{Samples} & \textbf{Users}&\textbf{Models}&\textbf{Charging\break Events}\\\hline
    
    Good & 3.3M&30K & 1.5K&180K \\\hline
    Over voltage& 1554&90 & 15 & 100\\\hline
    Overheat & 665 &165 & 40 & 200\\\hline
    \end{tabular}}
    
       \caption{Number of charging samples, corresponding users, device models, and charging events from  Carat dataset.}
     \label{tab:sample_stat}
    \end{center}
\end{table}

\begin{marginfigure}[1pc]
  \begin{minipage}{\marginparwidth}
  \vspace{-25mm}
  \subfigure[CC-CV \& DLC]{\label{fig:battery_voltage}\includegraphics[width=1.0\linewidth,height = 1.0\linewidth]{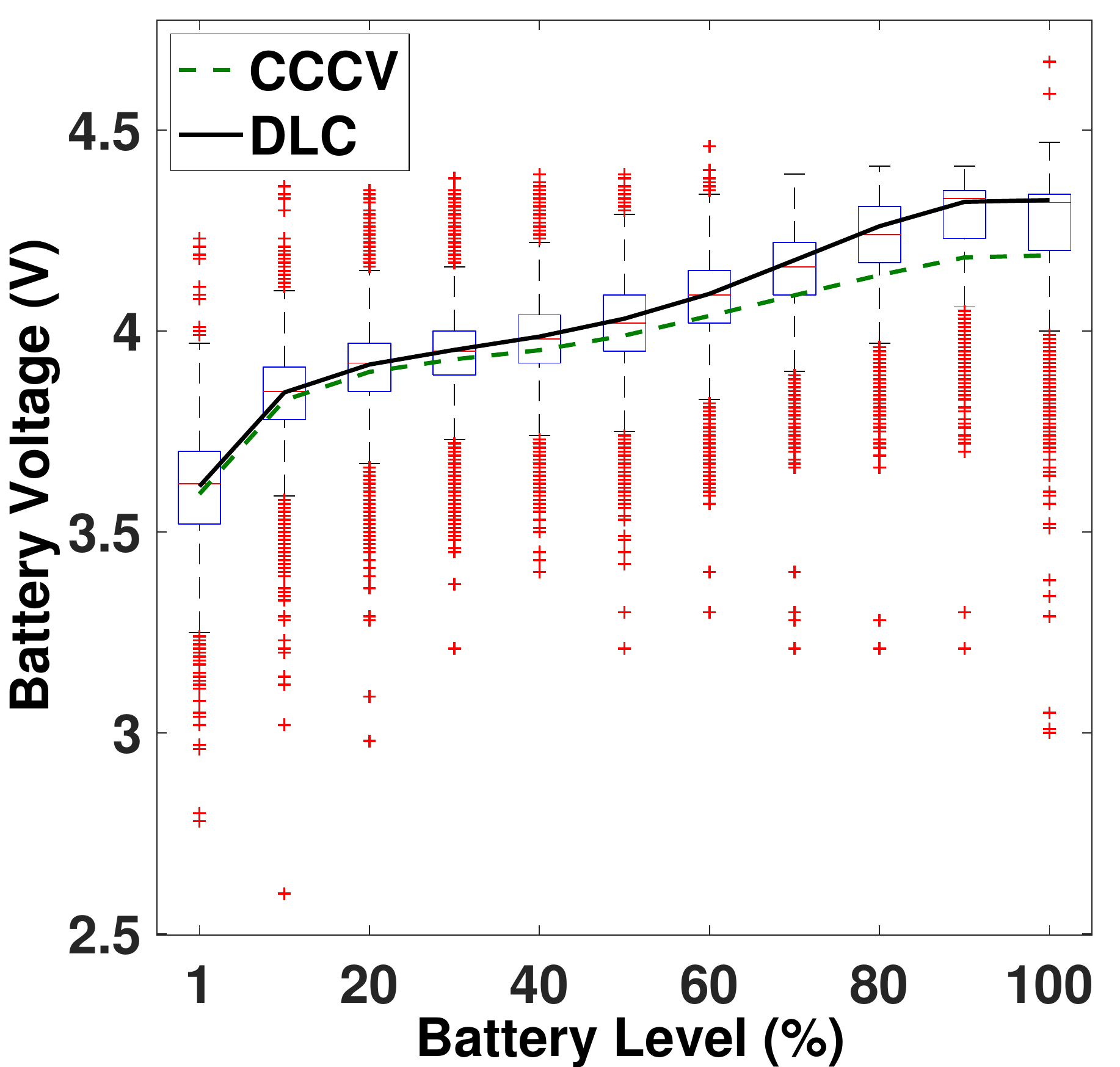}}
\subfigure[DLC \& Fast Charging]{\label{fig:charge_volt_compare}\includegraphics[width=1.02\linewidth,height = 1.1\linewidth]{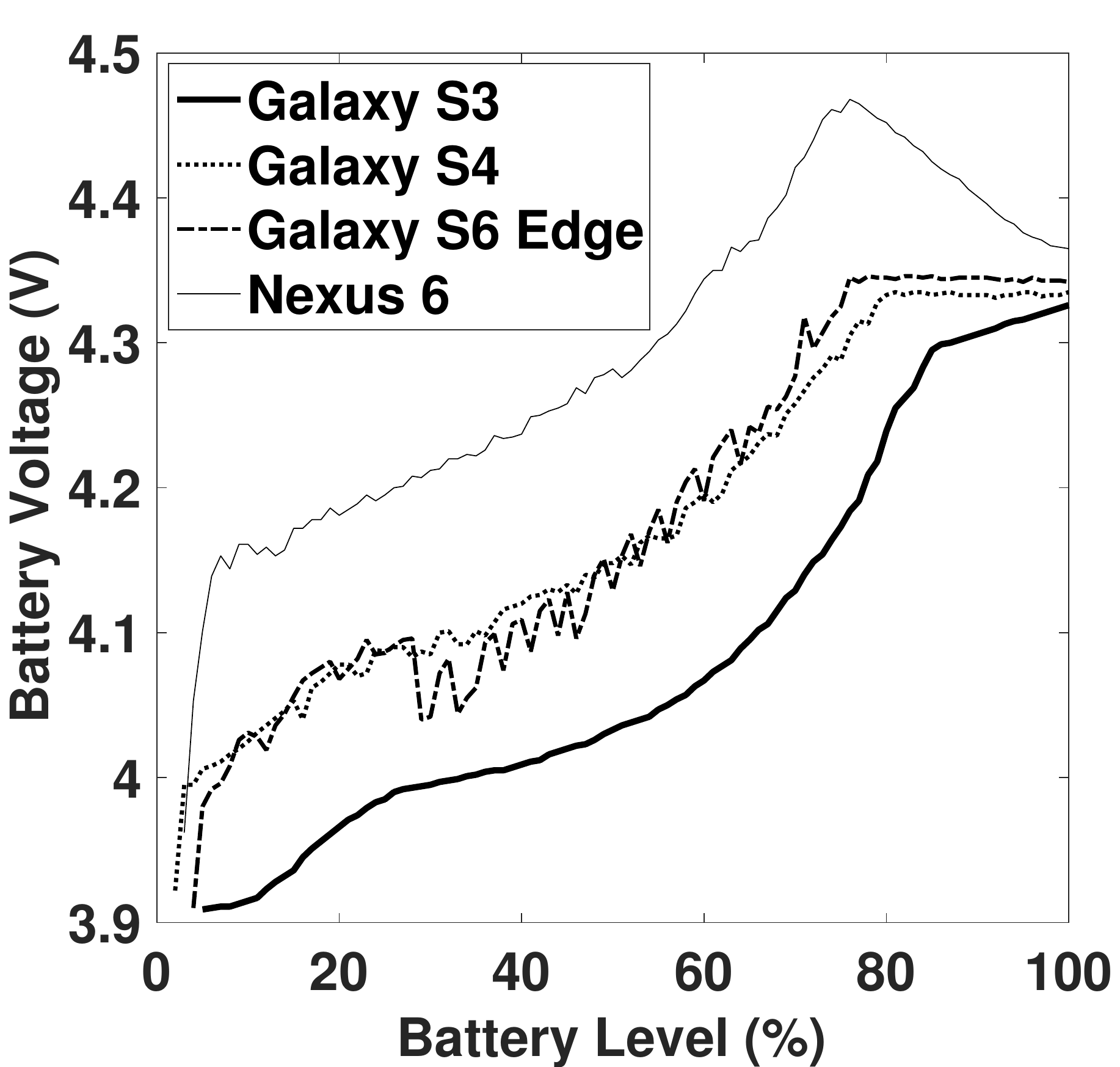}}
   \caption{Battery voltage curves while charging via AC with different charging techniques.}
\vspace{-3mm}
    \label{fig:battery_voltage_all}
  \end{minipage}
\end{marginfigure}

\section{Charging Techniques}
\label{sec:controller}
We first analyze  battery voltage behavior as charging proceeds and derive the charging techniques used by the smartphones. Next, we analyze the charging rates. In this section, we consider the charging events of good samples and further consider screen off samples from the events to reduce device usage bias in the analysis.

\subsection{Battery Voltage}
From the charging events, we construct  model specific charging voltage curves containing battery voltage information for every battery level update as shown in Figure~\ref{fig:battery_voltage}. The figure shows the \textit{initial} and \textit{final} voltages for each unique user. The initial voltage is the minimum voltage required to power up a device (when the battery level is 1\%) and final voltage is the maximum voltage when the battery is charged to 100\%. A battery should not be charged to more than this voltage.

The final voltages observed in the dataset broadly can be classified into two categories; 4.2$\pm$0.05 and 4.35$\pm$0.05 V. However, using CC-CV, a battery is charged to a maximum 4.2 V. It turns out that this voltage behavior is because of a different charging mechanism used by the devices. Although CC-CV is the well-known charging mechanism, some devices charge batteries to an extra 0.15 V to reduce charging hardware implementation complexity \cite{Thanh:2012}. This mechanism is also called Double Loop Control (DLC). We compare the final voltage of multiple devices of same model and find that 38\% of the devices use CC-CV and 59\% use DLC.

Figure~\ref{fig:battery_voltage} presents the distribution of voltage, and the relationship between battery voltage and battery level for the CC-CV and DLC methods. From the charging events, we take the median battery voltage for each battery level of the devices of two categories and plot them. It is shown that battery voltage increases almost linearly until the battery voltage reaches to the maximum, i.e., 4.2/4.35 V. Figure~\ref{fig:battery_voltage} further shows that the final voltage can be different for different users. Later, we demonstrate how this final voltage can be used to estimate capacity loss of the battery.

We next explore  model specific voltage curves. Figure~\ref{fig:charge_volt_compare} compares the median voltage curves of a number of devices. We notice that Galaxy S3 and S4 curves are similar as shown in Figure~\ref{fig:battery_voltage}, whereas the voltage curves of Galaxy S6 Edge, and Nexus 6 devices have unique characteristics. Unlike S3/S4, the battery voltage of Nexus 6 increases to a maximum 4.48 V and then the CV phase begins. On the other hand, the battery voltage of Edge 6 increases sharply until the battery is charged to 30\%. After that battery voltage increases and decreases alternatively until reaches to the maximum 4.35 V. This hints that Edge 6 applies Fast pulse charging~\cite{Thanh:2012}. In other words, the charging current alternates between two rates. Among 30K, only 3\% of the devices use Quick and Fast charging.

\vspace{-1mm}
\subsection{Charging Rate}
\label{subsec:three_two}


Charging algorithms may apply different charging currents to charge the batteries.  Android operating systems, however, do not expose charging rate or current through any API. We use the charging C rates estimated with Algorithm~\ref{alg:one}. In the earlier section, we have identified that batteries are charged in two phases, in general. In this section, we investigate the behavior of charging rates and present two additional charging mechanisms.

\begin{marginfigure}[1pc]
  \begin{minipage}{\marginparwidth}
  \vspace{-2mm}
  \subfigure[CC-phase rates]{\label{fig:charging_rates}\includegraphics[width=1.0\linewidth,height = 0.9\linewidth]{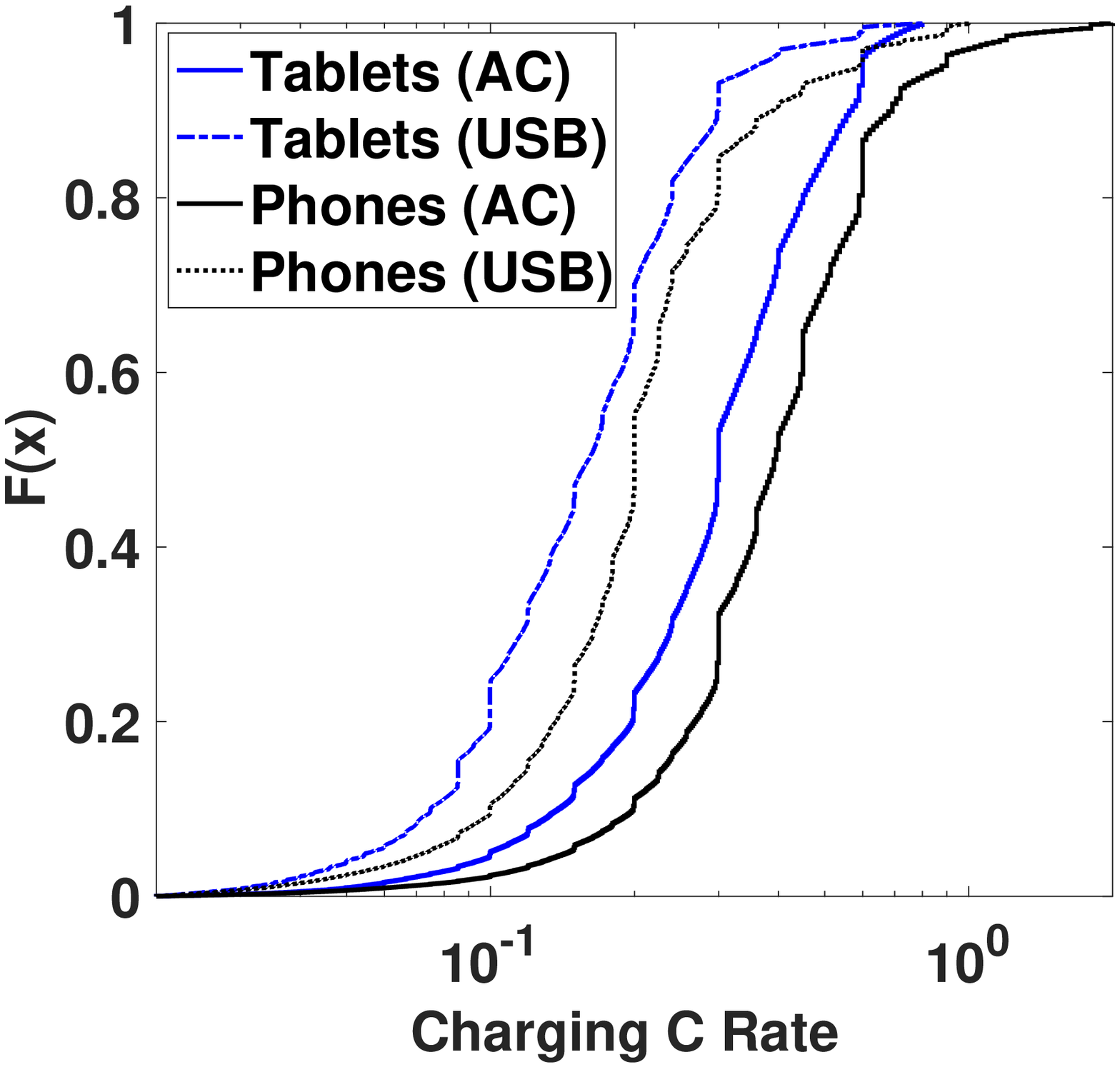}}
  \vspace{-3mm}
\subfigure[CV-phase rates]{\label{fig:cccv_dlc_rates}\includegraphics[width=1.02\linewidth,height = 0.9\linewidth]{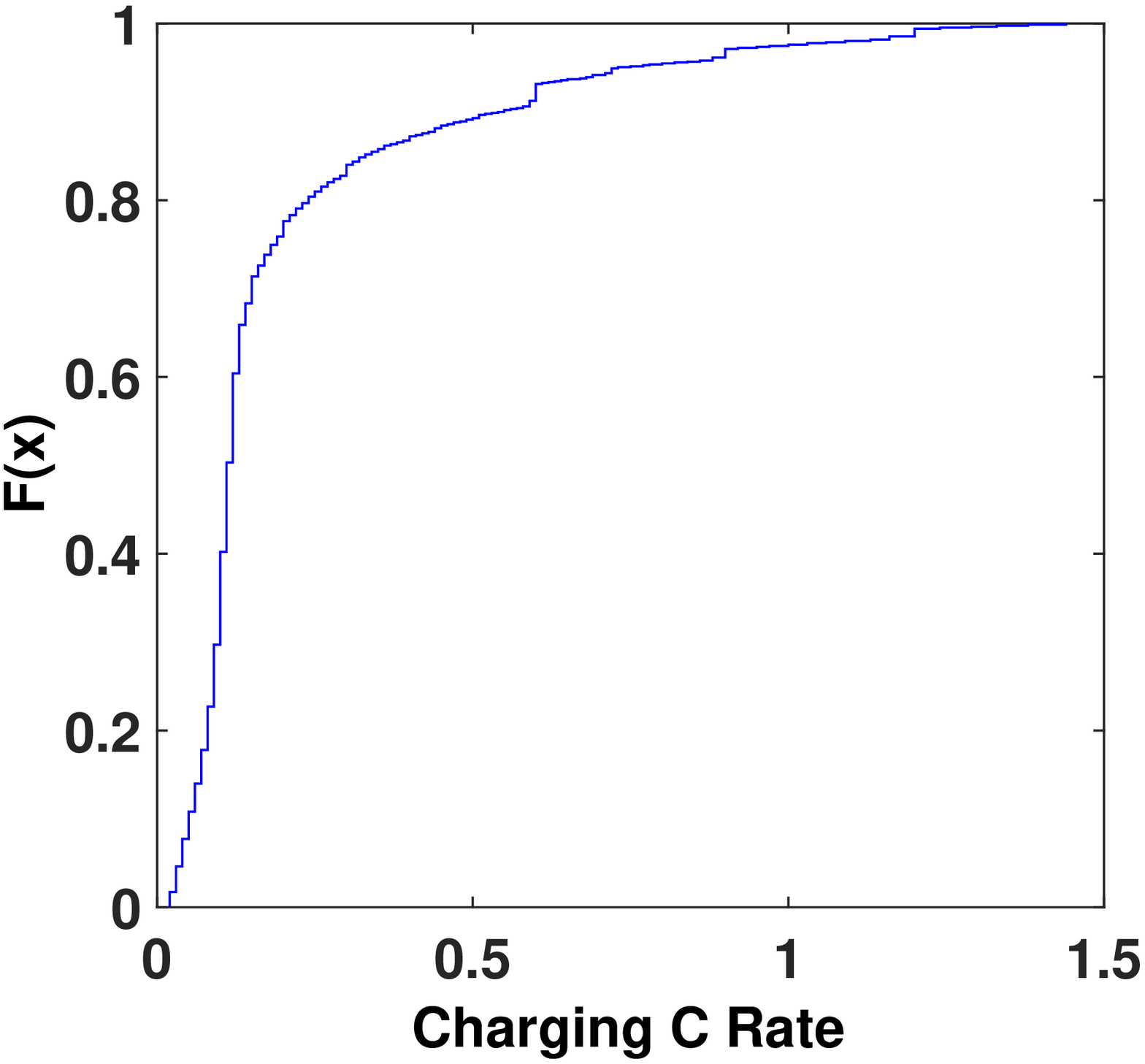}}
   \caption{Charging rates of the devices.}
    \label{fig:charging_all}
  \end{minipage}
  
  \includegraphics[width=1.0\textwidth,height = 0.7\linewidth]{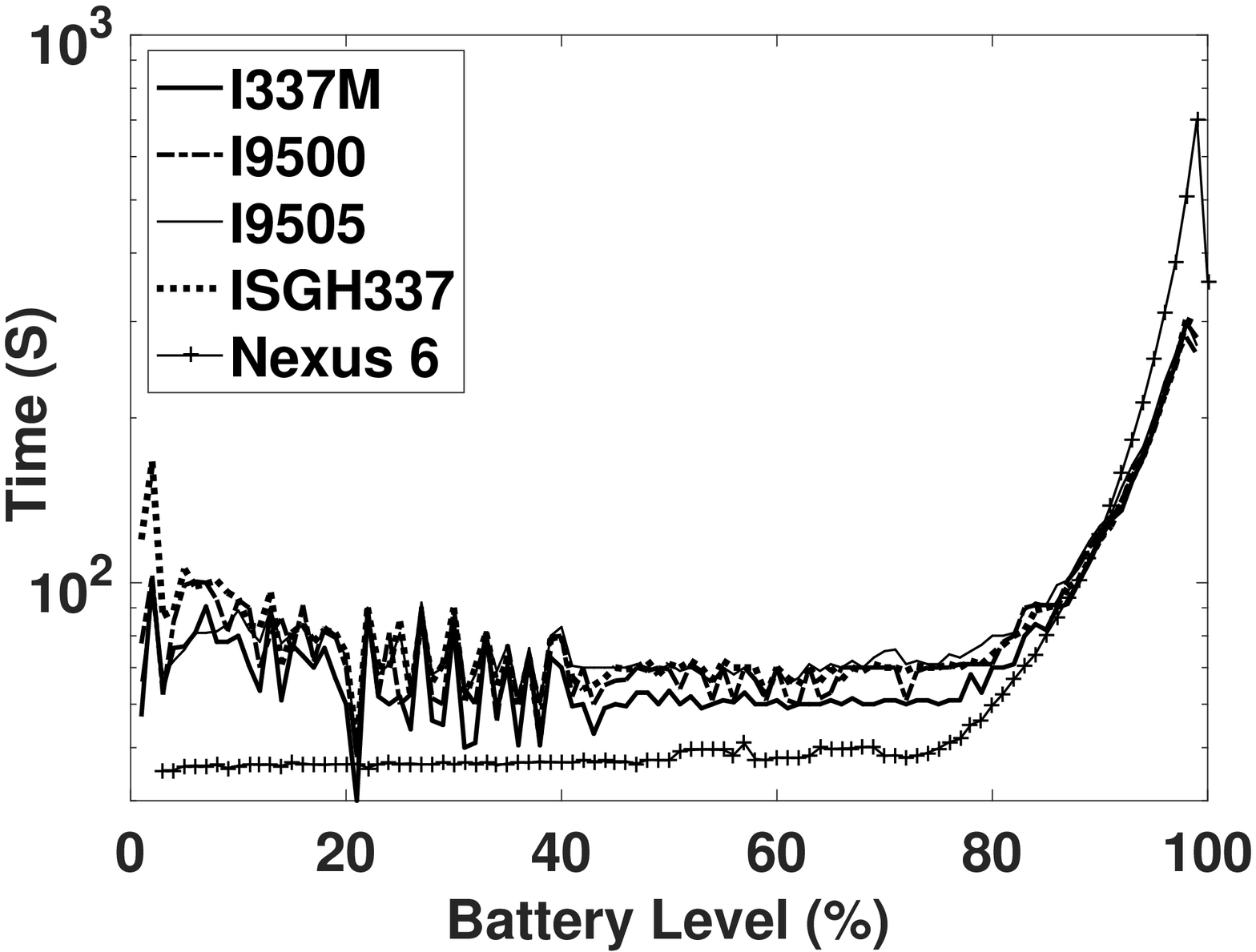}
    \caption{Charging time curves of five smartphone models.}
    \label{fig:onePercurve}

\end{marginfigure}

\noindent\textbf{CC-Phase: }
During the first phase, the batteries are charged to 50-90\% of total capacity, depending on the models. Figure~\ref{fig:charging_rates} illustrates the charging rates of 30K users. Mobile devices are charged mostly at rates smaller than 0.7C via AC and  higher than the rate USB provides. The rates are typically constant during the first phase and vary among different models. There are a number of takeaways from this figure.

\begin{itemize}
 
\item Figure~\ref{fig:charging_rates} shows that a few numbers of rates are higher than 1C for smartphones. By examining the charging events, we have  identified that 1\% of the devices use charging rates higher than 1C. These are the Fast charging enabled devices. 
 
 \item The tablets are charged at smaller rates than the smartphones, even when the AC main or wall chargers are used. One possible explanation is that tablets have larger batteries than the smartphones and their charging rates do not scale up according to the capacity.

\item Figure~\ref{fig:charging_rates} shows that  30\% and 20\% of the charging rates via USB are below 0.1C for the tablets and smartphones respectively. In the case of AC charging, 20\% of the rates are below 0.1C for the tablets, whereas such rates are less than 5\% for the smartphones. This is because a small number of devices apply the CV charging to charge the first 10\% and then the CC charging begins. This helps to restore the charge of deeply depleted cells inside the battery~\cite{Dearborn:2012}. Therefore, it took a longer time for an actual 1\% increment.

 \end{itemize}

\noindent\textbf{CV-Phase:}
Figure~\ref{fig:cccv_dlc_rates} shows the charging rates during the second phase of charging. Both  CC-CV and DLC methods trickle down the charging current gradually to less than or equal to 0.1C. Figure~\ref{fig:cccv_dlc_rates} also shows that charging rates can be around 1.0C. It turns out out that some device models use CC charging after the CV phase and this third phase begins after the battery is charged to 95\%  and the remaining 5\% is charged at a higher constant rate than the first phase.

\subsection{Summary}
Finally, other than CC-CV and DLC, we have identified two more variants of these two, one of them applies CV at the beginning and the other  uses CC at the end of charging. We have also identified two kinds of Fast charging technique. Their charging rates vary within 0.7-1.1C.  Nexus 6 takes 35 minutes, where as Galaxy 6 Edge takes 30 minutes to charge the first 50\% of the battery.



\section{SOC Estimation Techniques}
\label{sec:fuelgauge}
As mentioned in Section 2 that fuel gauge chips estimate SOC and smartphones basically employ either a voltage-based or Coulomb counter-based fuel gauge. The first kind depends on a number of voltage look up tables to estimate  SOC and the latter one uses current sense resistors to measure the charging/discharging current. It is difficult to identify the presence of such mechanisms without any explicit knowledge about the chipset model or name. In this section, we attempt to distinguish these two from their SOC reporting behavior.

From the charging time calculated in Section~2, we compute the median charging time curves of different models. Figure~\ref{fig:onePercurve} illustrates such curves for five models. We notice that SOC update times of Nexus 6 are almost constant until the battery is charged to 50\%. Given a device has a Coulomb-counter based fuel gauge, it can accurately measure SOC of the battery while charging. If the device is not utilized, the battery would receive the maximum constant charging current from the charger during  CC period. Therefore, the amount of time required to charge one percent should be equal for every SOC update within the CC-phase. The plot for Nexus 6 suggests that this model has Coulomb counter-based fuel gauge and we verified so.

On the other hand, the curves of \texttt{I337M, I9500, I9505}, and \texttt{ISGH337} are almost identical, however, they are different from Nexus 6. Their charging time vary at the same SOC. This hints that these devices use similar SOC estimation method. Later we find that these models are different variants of Samsung Galaxy S4 manufactured for different operators. We further verified that Galaxy S4 devices use voltage-based fuel gauges. Most of the devices in our dataset use voltage-based fuel gauges.

\section{Battery Properties}
\label{sec:battery}

We next investigate the battery properties, such as capacity, temperature behavior while charging, and health. Unlike the studies in other sections, we consider all the samples in this section.



\begin{marginfigure}[1pc]
  \begin{minipage}{\marginparwidth}
  \vspace{-30mm}
  \subfigure[Final Voltage vs. Capacity]{\label{fig:battery_capacity_rel}\includegraphics[width=1.0\linewidth,height = 1.0\linewidth]{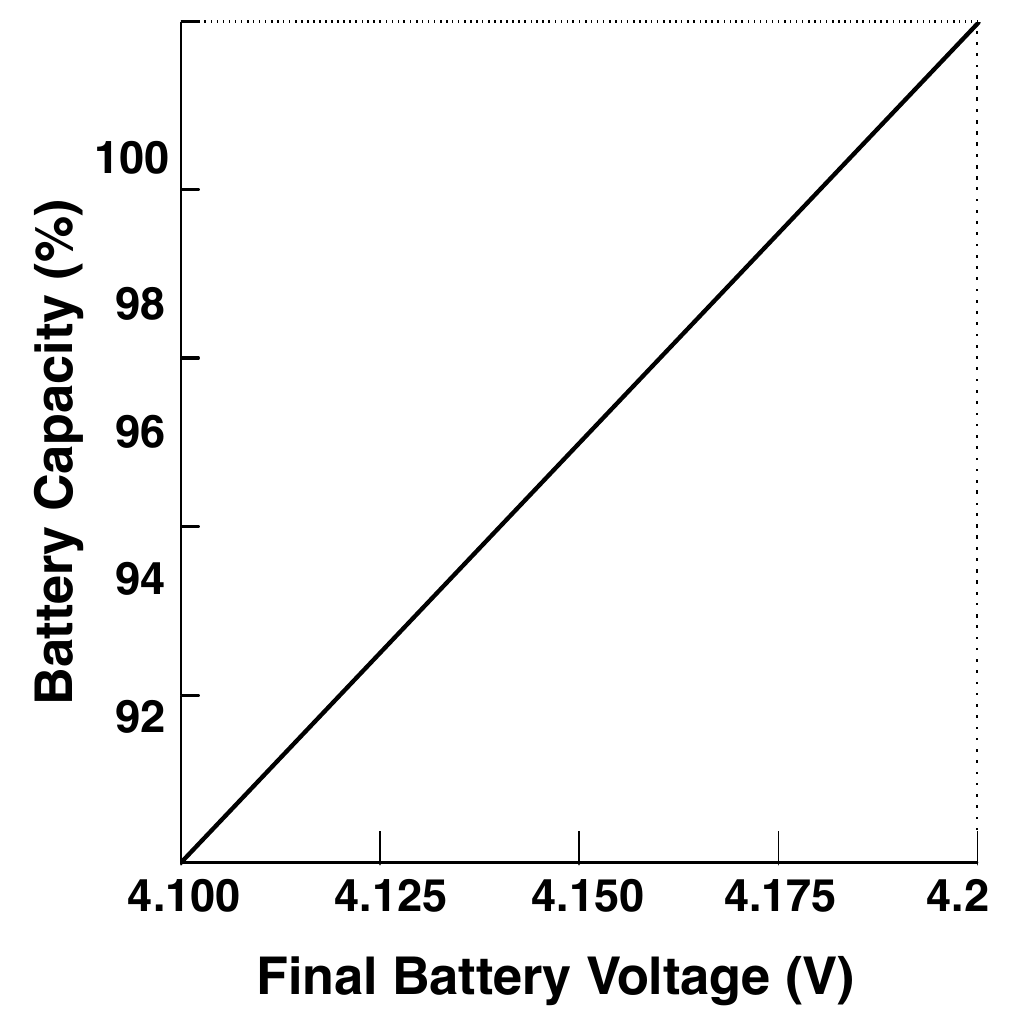}}
      \vspace{-3mm}
\subfigure[Battery Capacity]{\label{fig:battery_capacity_red}\includegraphics[width=1.0\linewidth,height = 0.9\linewidth]{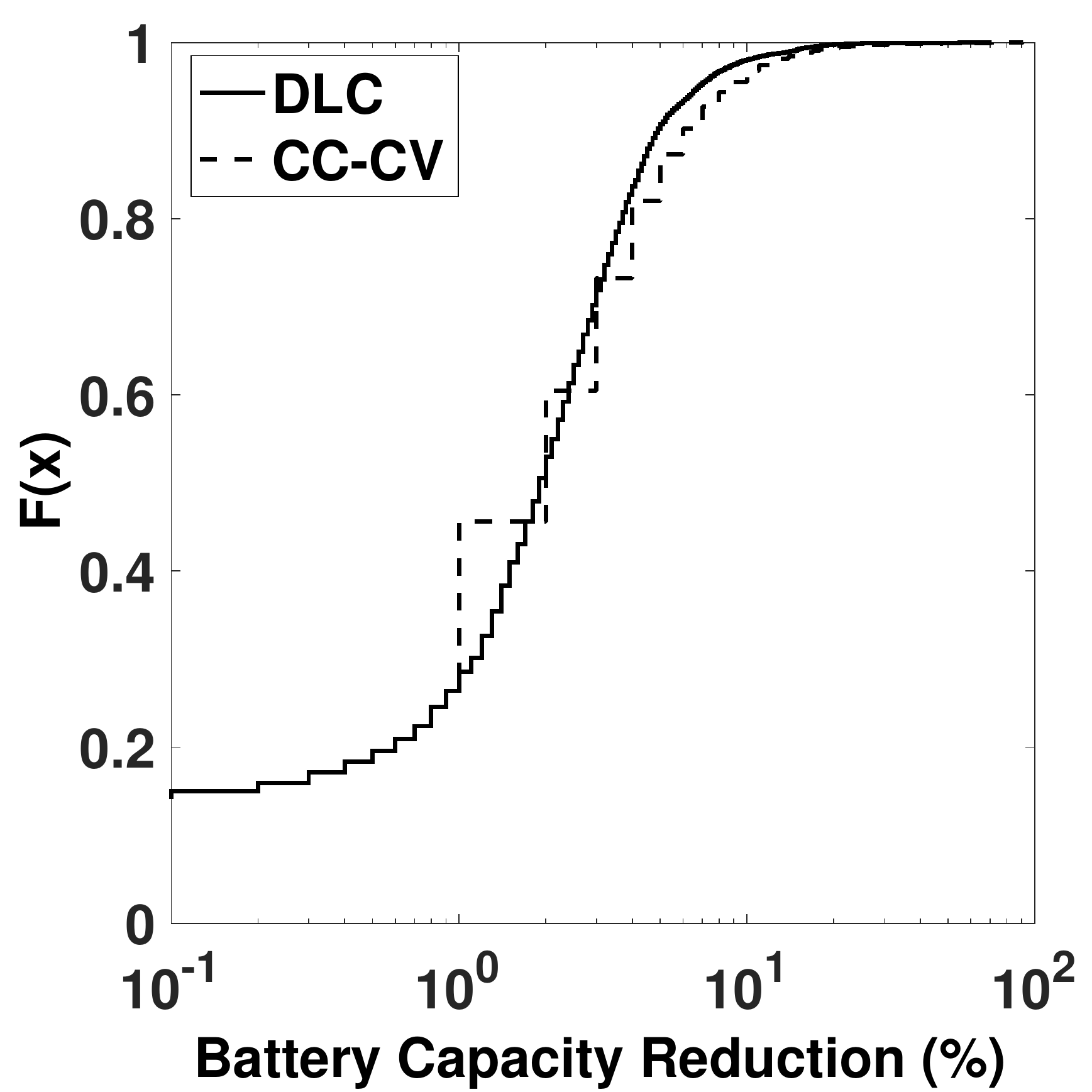}}
    \caption{Battery final voltage and capacity.}
    \label{fig:battery_capacity}
  \end{minipage}
  
  \includegraphics[width=1.0\textwidth,height = 0.7\linewidth]{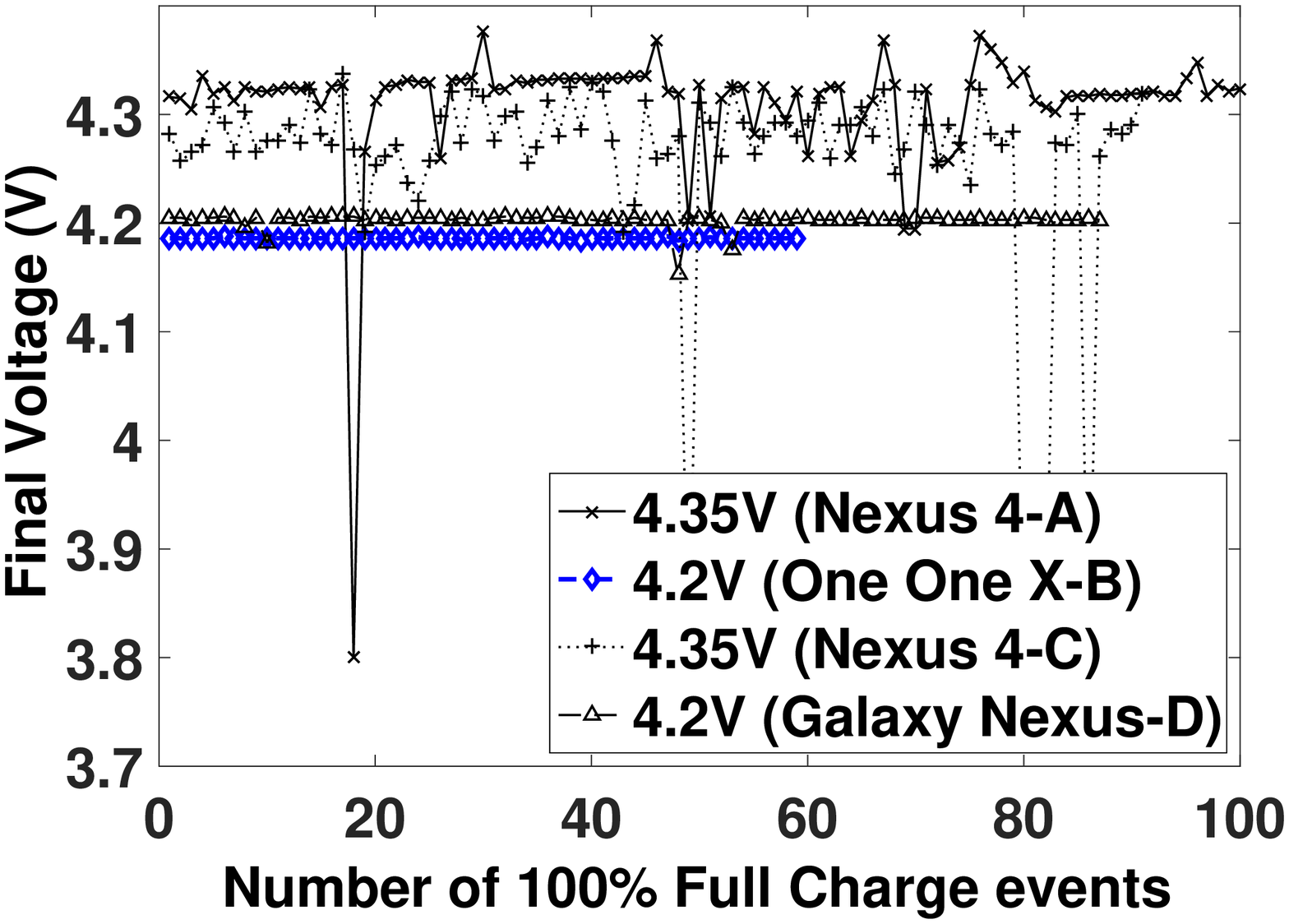}
    \caption{Final voltage over time.}
    \label{fig:charge_varone}

\end{marginfigure}

\subsection{Battery Capacity Loss}
The effect of charging with over voltage, and higher charging rate is degraded battery life. Choi and Lim~\cite{Choi2002130} studied the  performance of Lithium-Ion batteries by charging them with different charging rates and by limiting the final voltage. They showed that charging at higher rates (1.4C, 1.2C) and voltage (4.35 V) reduces battery capacity faster and significantly more than those with smaller magnitudes (1.0C, 4.2 V). However, it is not possible to estimate the effect of an individual variable without measurements. A recent study has shown that batteries with different capacity exhibit different voltage behavior while charging~\cite{Hoque:2015}. Although Android APIs do not expose the actual capacity in mAh, we can determine the relative remaining capacity or capacity loss from the final voltage. Figure~\ref{fig:battery_capacity_rel} shows that the relation between these two is linear~\cite{kestersection}. Every 10 mV reduction in the final voltage is equivalent to 1\% capacity loss. The capacity loss can be computed as 
\begin{equation}
Capacity_{Loss} (\%) = (V_{f} - V_{rf})/10,
\end{equation}

\noindent where the value of $V_{f}$ is 4.2 or 4.35 V and $V_{rf}$ is the reported final voltage in the sample. Another observation is that the final voltage for the DLC or 4.35V models fluctuates frequently (see Figure~\ref{fig:charge_varone}). Therefore, we take the average of all final voltages of an individual device and estimate the capacity loss. Figure~\ref{fig:battery_capacity_red} shows that 85\% of the devices have lost their capacity by 1-10\%. A number of devices have significant capacity loss. This information can be used as an input to the self-constructive power modeling approaches, such as PowerBooter~\cite{Zhang:2010} depends on relative battery capacity.

\subsection{Battery Temperature}

\begin{marginfigure}[1pc]
  \begin{minipage}{\marginparwidth}
  \vspace{-5mm}
  \subfigure[CC-CV \& DLC]{\label{fig:cccv_temperature}\includegraphics[width=1.0\linewidth,height = 1.0\linewidth]{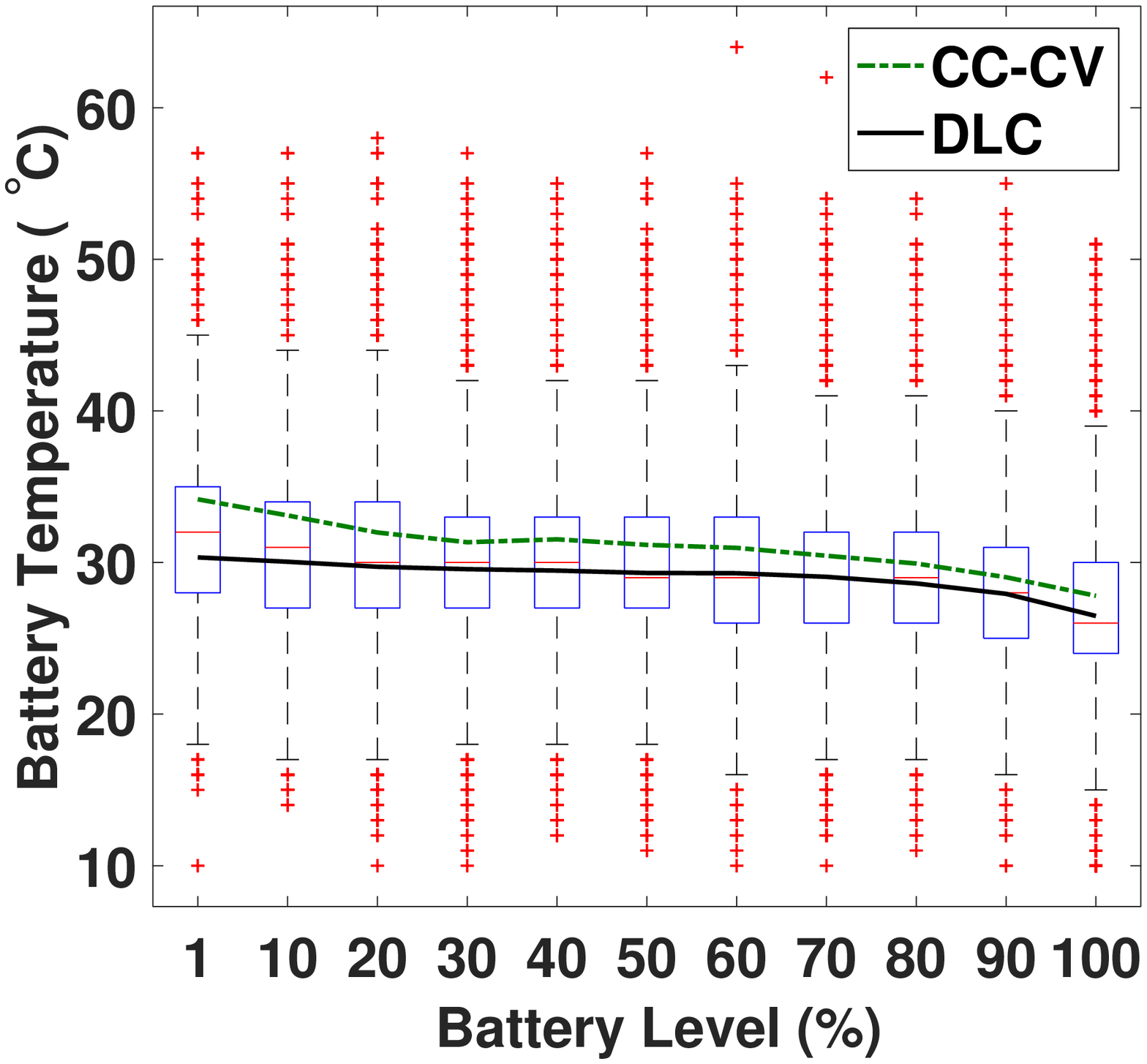}}
\subfigure[DLC \& Fast Charging]{\label{fig:fast_temperature}\includegraphics[width=1.04\linewidth,height = 1.0\linewidth]{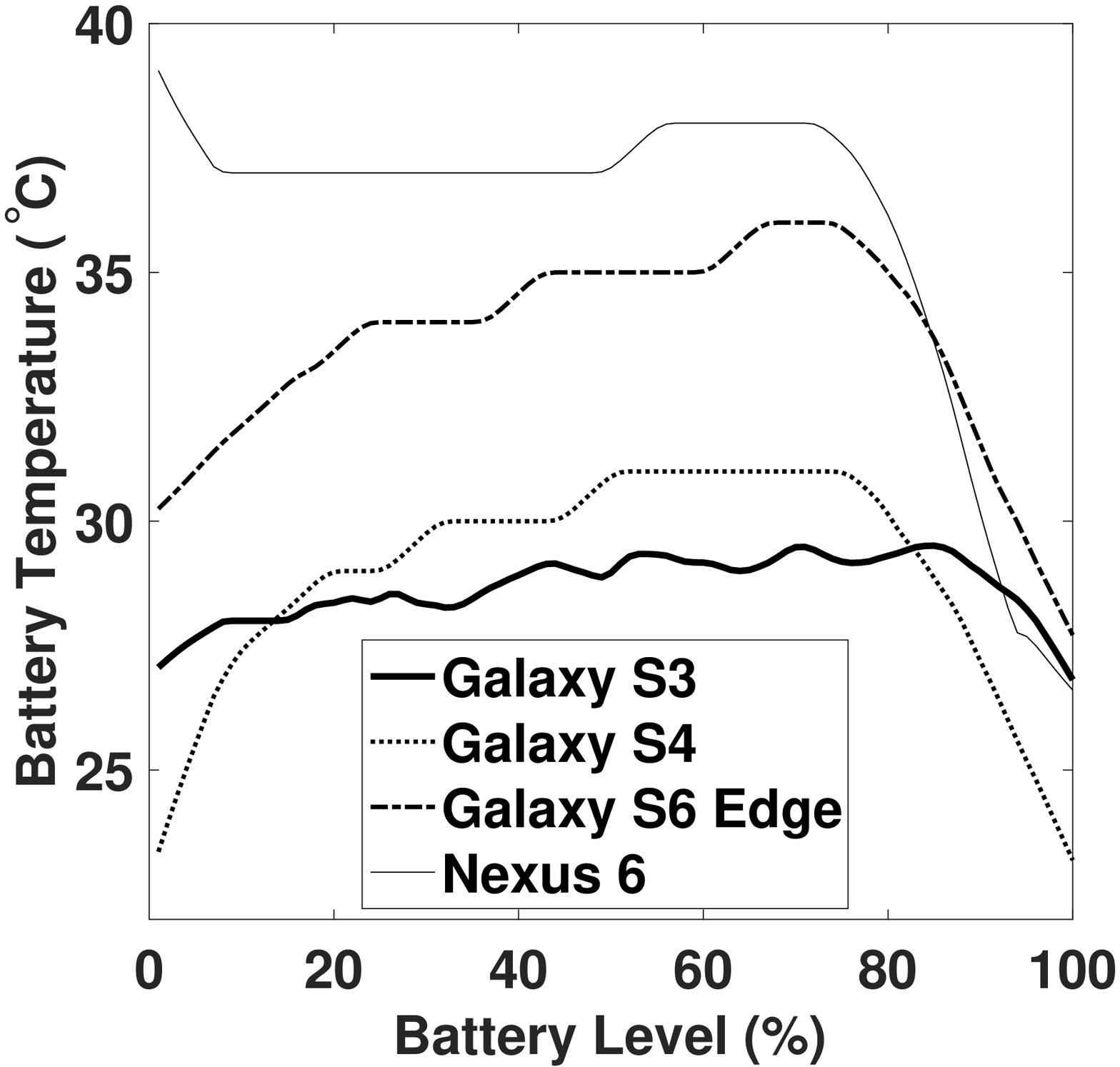}}
   \caption{{Battery temperature with different charging techniques via AC.}}
    \label{fig:battery_temperature}    
  \end{minipage}  
\end{marginfigure}

In order to understand the pattern of battery temperature while charging, we first group the battery temperature for each reported battery level and plot them in Figure~\ref{fig:cccv_temperature}. We also plot the median temperature for CC-CV and DLC  models. The battery temperature varies as the battery level increases. At the beginning, the temperature decreases till the battery is charged to 20\%. After that temperature slowly decreases or remains almost constant until the battery is charged to 70 or 80\%. However, at the end the temperature begins to decrease again. The obvious reason is very low current charging during the CV phase. Although both CC-CV and DLC exhibit similar temperature variation pattern, the average temperature of CC-CV models is higher than the DLC models. Figure~\ref{fig:fast_temperature} compares battery temperature between DLC and  Fast charging techniques. We notice that Fast charging increases battery temperature by 8-10$^{\circ}$C than that of the DLC model devices.

\subsection{Battery Health}
 Android battery APIs provide battery health information, such as good, over voltage, and overheat. Table~\ref{tab:battery_health_stat} shows the ranges of battery voltage and temperature for good, over voltage, and overheat samples. We notice that such ranges for over voltage and overheat samples also vary within the similar range as good samples. After examining these samples, we have found that over voltage samples do not give any hint about temperature and  similarly overheat samples do not hint whether over voltage results higher temperature.

\begin{margintable}[1pc]
  \begin{minipage}{\marginparwidth}
  {\footnotesize
    \begin{tabular}{lll}
      \hline
       \textbf{Heal.}  & \textbf{Volt.}& \textbf{Temp.}\\\hline
      Good & 3.2-4.4V&10-57$^\circ$C\\\hline
      Overheat & 3.2-4.32V &24-72$^\circ$C\\\hline
      Over volt. & 3.2-4.2V &16-44$^\circ$C\\\hline
   \end{tabular}}
      \caption{Battery health, voltage, and temperature information.}
    \label{tab:battery_health_stat}

  \end{minipage}
\end{margintable}

\vspace{-2mm}
\section{User Behavior in Charging}
\label{sec:user}
We consider all the charging events and the corresponding samples in this section as well. A charging event should contain one sample for each battery level or SOC update. Therefore, the number of samples for a specific battery level should be unique in an event. However, we have found more than one sample for a single battery level update in the form of SOC fluctuation (e.g. battery level = 5$\uparrow$6$\downarrow$5$\uparrow$6). Nevertheless, such SOC fluctuations are not uniformly distributed, rather left skewed with respect to the battery level.

2\% of the charging events reside at the tail of the distribution, which contain fluctuation between two consecutive levels. The screen status of the corresponding samples suggests that the devices were being actively used. Therefore, it took a longer time for an actual 1\% increment.

Other than charging and actively using their devices at the same time, users may keep their devices connected with the chargers even when the batteries are completely charged. From the dataset, we have identified  3\% of such charging events. The duration  of such events can be a few to thousands of seconds. In this case, the phone stops charging the battery and begins recharging whenever 1-2\% has been discharged. However, we have measured that the extra energy spent during a over night charging for 10 hours can be used to charge a iPhone 6 to its full capacity (1810mAh).

\section{Conclusions and Future Work}
\label{sec:conclusion}

In this study based on data gathered from in-the-wild devices, we have shown that a few thousand devices use inefficient charging mechanisms that can significantly reduce battery life. We have found that 2\% of the devices charge their batteries well above the maximum battery voltage. This charging method deteriorates battery capacity faster than normal. There has been very active discussion in various online forums identifying battery SOC anomalies and such SOC error is due to the capacity loss~\cite{Hoque:2015}. A small number of devices had a charging current higher than 1.0C, which also degrades battery performance quickly. We have also observed that 85\% of the devices suffered from 1-10\% capacity loss. Moreover, user behavior and interaction with the device during charging also contribute to energy waste. Our future research includes investigating the performance of different charging algorithms with a larger dataset and developing a battery analytics API based on Spark so that users and vendors can investigate the performance of their batteries and power management techniques.

\section{Acknowledgements}
This work was funded by the Academy of Finland CUBIC project with grant number 277498.
\balance{}


\begin{thebibliography}{00}


\ifx \showCODEN    \undefined \def \showCODEN     #1{\unskip}     \fi
\ifx \showDOI      \undefined \def \showDOI       #1{{\tt DOI:}\penalty0{#1}\ }
  \fi
\ifx \showISBNx    \undefined \def \showISBNx     #1{\unskip}     \fi
\ifx \showISBNxiii \undefined \def \showISBNxiii  #1{\unskip}     \fi
\ifx \showISSN     \undefined \def \showISSN      #1{\unskip}     \fi
\ifx \showLCCN     \undefined \def \showLCCN      #1{\unskip}     \fi
\ifx \shownote     \undefined \def \shownote      #1{#1}          \fi
\ifx \showarticletitle \undefined \def \showarticletitle #1{#1}   \fi
\ifx \showURL      \undefined \def \showURL       #1{#1}          \fi

\bibitem{Banerjee:2007}
{Nilanjan Banerjee}, {Ahmad Rahmati}, {Mark~D. Corner}, {Sami Rollins}, {and}
  {Lin Zhong}. 2007.
\newblock \showarticletitle{Users and Batteries: Interactions and Adaptive
  Energy Management in Mobile Systems}. In {\em Proceedings of the 9th
  International Conference on Ubiquitous Computing} {\em (UbiComp '07)}.
  Berlin, Heidelberg, 217--234.
\newblock
\showISBNx{978-3-540-74852-6}


\bibitem{Choi2002130}
{Soo~Seok Choi} {and} {Hong~S Lim}. 2002.
\newblock \showarticletitle{Factors that affect cycle-life and possible
  degradation mechanisms of a Li-ion cell based on LiCoO2}.
\newblock {\em Journal of Power Sources\/} {111}, 1 (2002), 130 -- 136.
\newblock
\showISSN{0378-7753}


\bibitem{Dearborn:2012}
{Scott Dearborn}. 2005.
\newblock {\em Charging Lithium-Ion batteries for Maximum Run Times}.
\newblock {T}echnical {R}eport.
\newblock
\newblock
\shownote{\url{http://powerelectronics.com/site-files/powerelectronics.com/files/archive/powerelectronics.com/mag/504PET23.pdf}.}


\bibitem{Ferreira:2011}
{Denzil Ferreira}, {Anind~K. Dey}, {and} {Vassilis Kostakos}. 2011.
\newblock \showarticletitle{Understanding Human-smartphone Concerns: A Study of
  Battery Life}. In {\em Proceedings of the 9th International Conference on
  Pervasive Computing} {\em (Pervasive'11)}. Berlin, Heidelberg, 19--33.
\newblock
\showISBNx{978-3-642-21725-8}


\bibitem{hoquecsur2015}
{Mohammad~Ashraful Hoque}, {Matti Siekkinen}, {Kashif~Nizam Khan}, {Yu Xiao},
  {and} {Sasu Tarkoma}. 2015.
\newblock \showarticletitle{Modeling, Profiling, and Debugging the Energy
  Consumption of Mobile Devices}.
\newblock {\em ACM Comput. Surv.\/} {48}, 3, Article 39 (Dec. 2015), 40 pages.
\newblock
\showISSN{0360-0300}


\bibitem{Hoque:2015}
{Mohammad~A. Hoque} {and} {Sasu Tarkoma}. 2015.
\newblock \showarticletitle{Sudden Drop in the Battery Level?: Understanding
  Smartphone State of Charge Anomaly}. In {\em Proceedings of the Workshop on
  Power-Aware Computing and Systems} {\em (HotPower '15)}. ACM, New York, NY,
  USA, 26--30.
\newblock
\showISBNx{978-1-4503-3946-9}


\bibitem{kestersection}
{Walt Kester} {and} {Joe Buxton}. 1996.
\newblock \showarticletitle{SECTION 5 BATTERY CHARGERS: Practical Design
  Techniques for Power and Thermal Management. s.l. : Analog Devices.}
\newblock  (1996).
\newblock


\bibitem{Oliner:2013}
{Adam~J. Oliner}, {Anand~P. Iyer}, {Ion Stoica}, {Eemil Lagerspetz}, {and}
  {Sasu Tarkoma}. 2013.
\newblock \showarticletitle{Carat: Collaborative Energy Diagnosis for Mobile
  Devices}. In {\em Proceedings of the 11th ACM Conference on Embedded
  Networked Sensor Systems} {\em (SenSys '13)}. ACM, New York, NY, USA, Article
  10, 14 pages.
\newblock
\showISBNx{978-1-4503-2027-6}


\bibitem{qulcomm}
{{Qualcomm}}. 2016.
\newblock {Quick} {Charge}.
\newblock   (2016).
\newblock
\newblock
\shownote{\url{https://www.qualcomm.com/products/snapdragon/quick-charge}.}


\bibitem{Rezvanizaniani2014110}
{Seyed~Mohammad Rezvanizaniani}, {Zongchang Liu}, {Yan Chen}, {and} {Jay Lee}.
  2014.
\newblock \showarticletitle{Review and recent advances in battery health
  monitoring and prognostics technologies for electric vehicle (EV) safety and
  mobility}.
\newblock {\em Journal of Power Sources\/} {256}, 0 (2014), 110 -- 124.
\newblock
\showISSN{0378-7753}


\bibitem{6157576}
{Narseo Vallina-Rodriguez} {and} {Jon Crowcroft}. 2013.
\newblock \showarticletitle{Energy Management Techniques in Modern Mobile
  Handsets}.
\newblock {\em Communications Surveys Tutorials, IEEE\/} {15}, 1 (2013),
  179--198.
\newblock
\showISSN{1553-877X}


\bibitem{Thanh:2012}
{Thanh~Tu Vo}, {Weixiang Shen}, {and} {A. Kapoor}. 2012.
\newblock \showarticletitle{Experimental comparison of charging algorithms for
  a lithium-ion battery}. In {\em IPEC, 2012 Conference on Power Energy}.
  207--212.
\newblock


\bibitem{spark}
{Matei Zaharia}, {Mosharaf Chowdhury}, {Tathagata Das}, {Ankur Dave}, {Justin
  Ma}, {Murphy McCauly}, {Michael~J. Franklin}, {Scott Shenker}, {and} {Ion
  Stoica}. 2012.
\newblock \showarticletitle{Resilient Distributed Datasets: A Fault-Tolerant
  Abstraction for In-Memory Cluster Computing}. In {\em Presented as part of
  the 9th USENIX Symposium on Networked Systems Design and Implementation (NSDI
  12)}. USENIX, San Jose, CA, 15--28.
\newblock
\showISBNx{978-931971-92-8}


\bibitem{Zhang:2010}
{Lide Zhang}, {Birjodh Tiwana}, {Zhiyun Qian}, {Zhaoguang Wang}, {Robert~P.
  Dick}, {Zhuoqing~Morley Mao}, {and} {Lei Yang}. 2010.
\newblock \showarticletitle{Accurate Online Power Estimation and Automatic
  Battery Behavior Based Power Model Generation for Smartphones}. In {\em
  Proceedings of the Eighth IEEE/ACM/IFIP International Conference on
  Hardware/Software Codesign and System Synthesis} {\em (CODES/ISSS '10)}. ACM,
  New York, NY, USA, 105--114.
\newblock
\showISBNx{978-1-60558-905-3}


\end{thebibliography}
\end{document}